\documentstyle[12pt,twoside,epsfig,espcrc1]{article}

\newcommand{\AmS}{{\protect\the\textfont2
  A\kern-.1667em\lower.5ex\hbox{M}\kern-.125emS}}

\begin{document}

\title{Asymmetry Effects on Nuclear Fragmentation} 

\author{M.Di Toro$^a$, V.Baran$^{a,b}$, M.Colonna$^a$, S.Maccarone$^a$,
 M.Zielinska-Pfabe'$^c$ and H.H.Wolter$^d$}

\address{ Laboratorio Nazionale del Sud, Via S. Sofia 44,
I-95123 Catania, Italy\\   and University of Catania}
\address{ NIPNE, Bucharest, Romania}
\address{ Smith College, Northampton, USA}
\address{ Sektion Physik, University of Munich, Germany}

%

\maketitle

\begin{abstract}
We show the possibility of extracting important information
on the symmetry term of the Equation of State ($EOS$) directly
from multifragmentation reactions using stable isotopes with different 
charge asymmetries. We study n-rich and n-poor $Sn + Sn$ collisions
at $50AMeV$ using a new stochastic transport approach with all
isospin effects suitably accounted for. For central collisions
a chemical component in the spinodal instabilities is clearly seen.
This effect is reduced in the neck fragmentation observed
for semiperipheral collisions, pointing to a different nature
of the instability. In spite of the low asymmetry tested
with stable isotopes the results are showing an interesting
and promising dependence on the stiffness of the symmetry term,
with an indication towards an increase of the repulsion
above normal density.
\end{abstract}

\section{Introduction}

Our starting point is that the key question in the physics of
unstable nuclei is the knowledge of the $EOS$ for
asymmetric nuclear matter away from normal conditions.
We remark the effect of the symmetry term at low densities
on the neutron skin structure, while the knowledge in
high densities region is crucial for supernovae dynamics
and neutron star cooling \cite{bom94,irv78,dtt92,pet93,urp95}.

Effective interactions are obviously tuned to symmetry properties
around normal conditions and any extrapolation can be quite
dangerous. Microscopic approaches based on realistic $NN$
interactions, Brueckner or variational schemes, or on effective
field theories show a quite large variety of predictions,
see Fig.1.

In the reaction dynamics with intermediate energy radioactive beams
we can probe highly asymmetric nuclear matter in compressed as well as
dilute phases: the aim of this paper is to show that fragmentation
events have new features due to isospin effects and that some
observables are particularly sensitive to the symmetry
term of the $EOS$.

\begin{figure}[ht]
\begin{center}
\mbox{\psfig{file=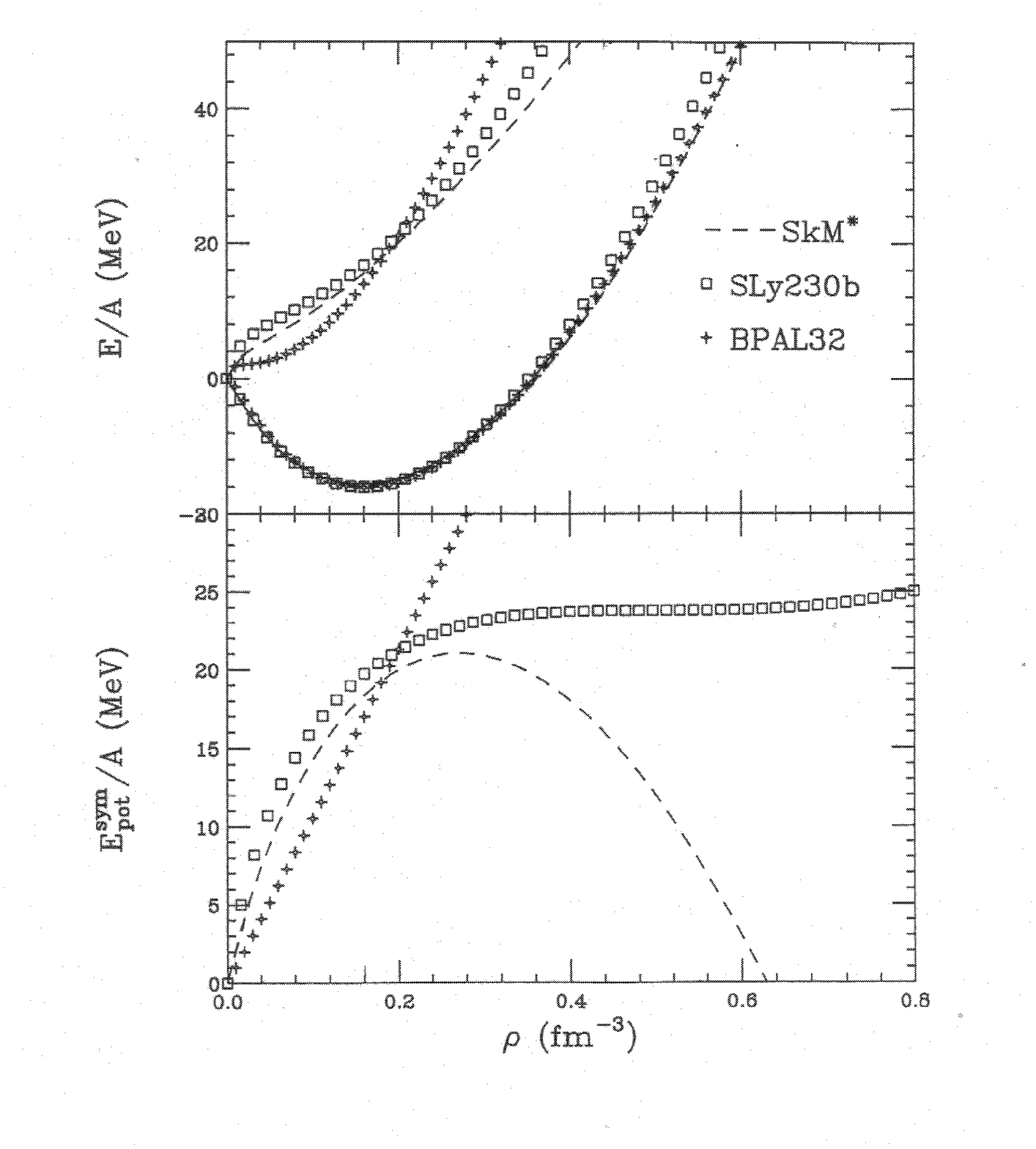,height=7.truecm}}
\end{center}
{\footnotesize
\begin{quotation}
\noindent
{\bf Fig.1 - } $EOS$ for various effective forces, $SKM^*$
\cite{kri80}(dashed), $SLy230b$ \cite{lyon97}(squares) and
$BPAL32$ \cite{bom94}(crosses). Top: neutron matter (up), symmetric matter
(down); Bottom: potential symmetry term. 
\end{quotation}}
\end{figure}

For fragmentation mechanisms a new qualitative feature is expected:
the onset of chemical effects on spinodal 
instabilities \cite{mue95,bao197,cdl98}, due to the intrinsic coupling
between isoscalar and isovector modes. The formation of a very neutron rich
gas vs. an almost symmetric liquid phase is expected for neutron excess
systems, in a dynamical non-equilibrium mechanism on short time scales
\cite{cdl98,bar98,dit98}. This effect, which appears very much reduced 
in a
statistical multifragmentation calculation \cite{dyst99}, will show
up in a production of more stable primary intermediate mass
fragments and in enhanced yields for neutron rich light isobars.

All these predictions were actually based on linear response
approaches \cite{cdl98,bar98}, nuclear matter dynamics in a box 
\cite{bar98,dit98} or on the evolution of a suitably prepared
excited source \cite{dyst99}. In this paper we will present the
first results of a fully "ab initio" calculation of fragmentation
reactions for systems with different charge asymmetry.

\section{"Ab initio" simulations of Sn + Sn reactions}

A new code for the solution of microscopic transport equations,
the {\it Stochastic Iso-BNV},
has been written where asymmetry effects are
suitably accounted for and the dynamics of fluctuations is
 included \cite{fab98,flu98}. A density dependent symmetry term is
used also in the ground state construction of the initial
conditions. Isospin effects on nucleon cross section \cite{lim9394}
and
Pauli blocking are consistently evaluated. 
In order to simplify the analysis of the most sensitive observables
to isospin effects we have chosen a Skyrme force with
the same $soft~EOS$ for symmetric Nuclear Matter ($NM$) ($K=201MeV$)
 and with two different choices
for the density dependence of the symmetry term, see Fig.1(bottom),
{\it asy-stiff} (like in $BPAL32$, see also \cite{qhd}) 
and {\it asy-soft} (like in $SKM^*$).
In this way we force the symmetric
part of the $EOS$ to be exactly the same in order to 
disentangle dynamical symmetry term effects.
We will show that the reaction mechanism is sensitive to the
different behaviours, although with stable nuclei we will limit the
possible asymmetries and moreover we will
certainly not
reach high compression regions. 

We have studied the $50AMeV$ collisions of the systems 
$^{124}Sn+^{124}Sn$ and $^{112}Sn+^{112}Sn$, where new data
are under analysis at $NSCL-MSU$ \cite{betty}.
We will comment $100$ events generated in semi-central ($b=2fm$) 
and semi-peripheral ($b=6fm$) reactions. In Fig.2 we show
a typical impact parameter evolution of the density plot 
(projected on the reaction plane) for one event
(neutron rich case, asy-stiff $EOS$).

\begin{figure}[ht]
\begin{center}
\mbox{\psfig{file=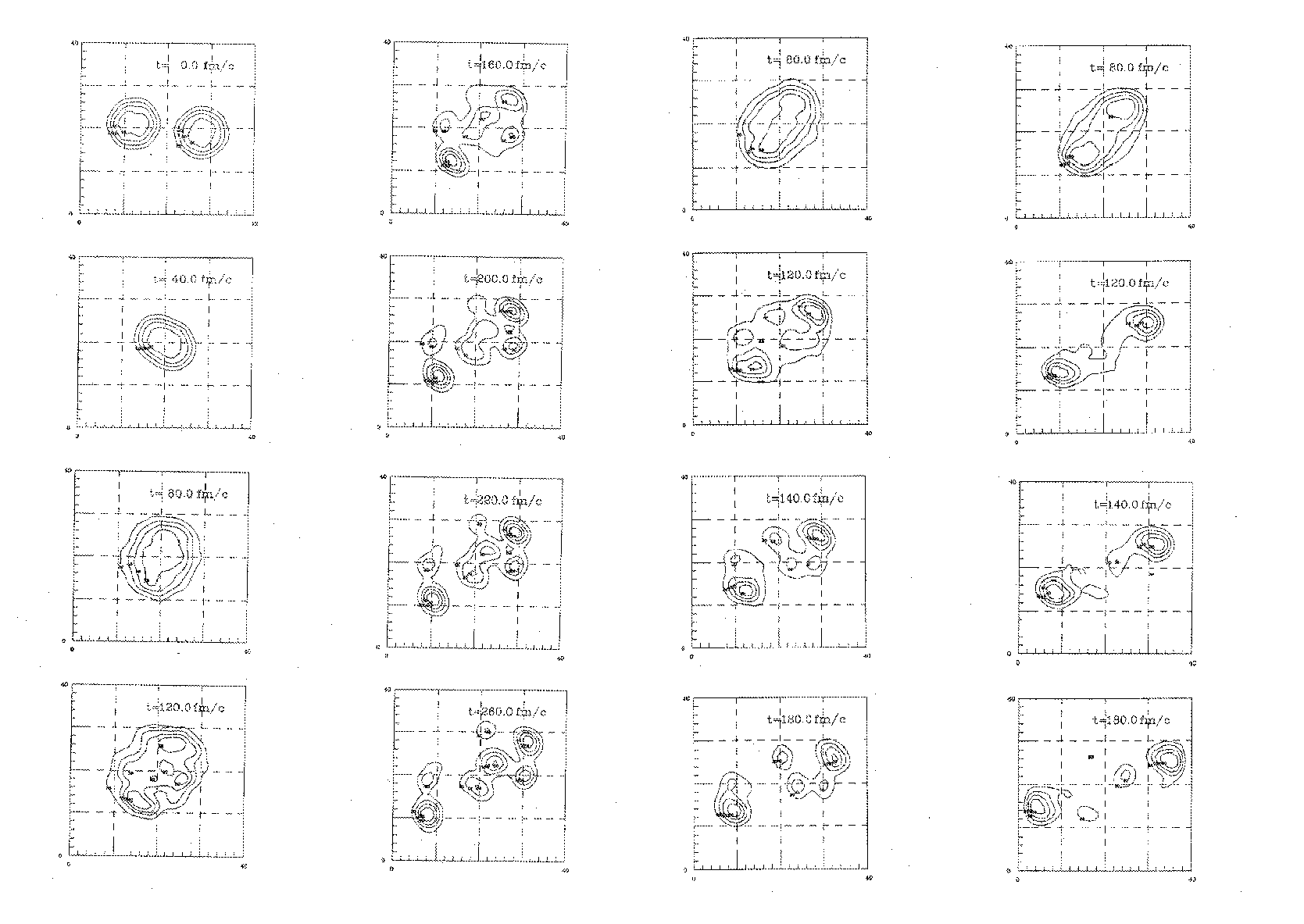,height=8.truecm}}
\end{center}
{\footnotesize
\begin{quotation}
\noindent
{\bf Fig.2 - } $^{124}Sn+^{124}Sn$ $50AMeV$: time
evolution of the nucleon density projected on the reaction plane.
First two columns: $b=2fm$ collision, approaching, compression and 
separation phases. Third column: $b=4fm$ , fourth column: $b=6fm$,
separation phase up to the {\it freeze-out}.
\end{quotation}}
\end{figure}

We remark: i) In the cluster formation we see a quite clear transition from 
{\it bulk}  \cite{col94,cdg94,col98} to {\it neck} \cite{col95,inri97,dur98}
instabilities. ii) The {\it "Freeze-Out times"}, when the nuclear
interaction among clusters disappears, are decreasing with impact parameter. 
These two dynamical
effects will influence the isospin content of the produced primary
fragments, as shown later.

A detailed analysis of the results from $100$ events for the same system
({\it asy-stiff} calculation) is shown in Fig.3 (central, $b=2fm$)
and Fig.4 (peripheral, $b=6fm$). Each figure is organized in this way:

Top row, time evolution of: (a) {\it Mass} in the
liquid (up) and gas (down) phase; (b) {\it Asymmetry} $I=(N-Z)/(N+Z)$ in
the gas "central" (solid line and squares), gas total (dashed+squares), 
liquid
"central" (solid+circles) and clusters (stars). 
"Central" means in a cubic box of side $20fm$
around the c.m..
The horizontal line shows the initial
average asymmetry; (c) {\it Mean Fragment Multiplicity} $Z \geq 3$.
The saturation of this curve defines the freeze-out configuration,
as we can also check from the density plots like in Fig.2.

Bottom row, properties of the "primary" fragments in the
{\it Freeze-Out Configuration}: (d) {\it Charge Distribution}, 
 (e) {\it Asymmetry Distribution} and
(f) {\it Fragment Multiplicity Distribution} (normalized to $1$).

\begin{figure}
\begin{center}
\mbox{\psfig{file=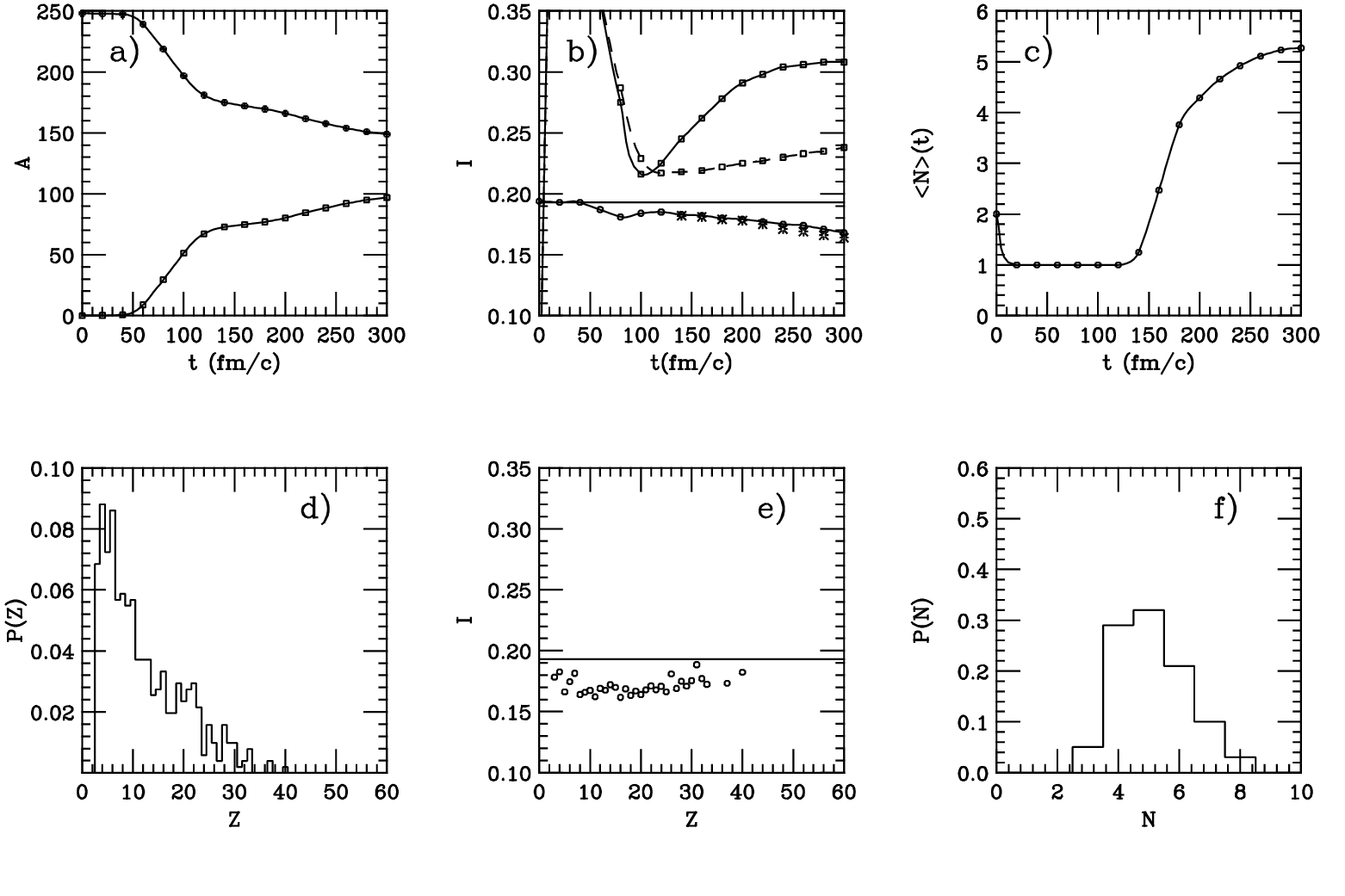,height=5.6truecm}}
\end{center}
{\footnotesize
\begin{quotation}
\noindent
{\bf Fig.3 - } $^{124}Sn+^{124}Sn$ $50AMeV$ $b=2fm$ collisions: time
evolution and freeze-out properties.
See text. {\it ASY-STIFF} $EOS$.
\end{quotation}}
\end{figure}

\begin{figure}
\begin{center}
\mbox{\psfig{file=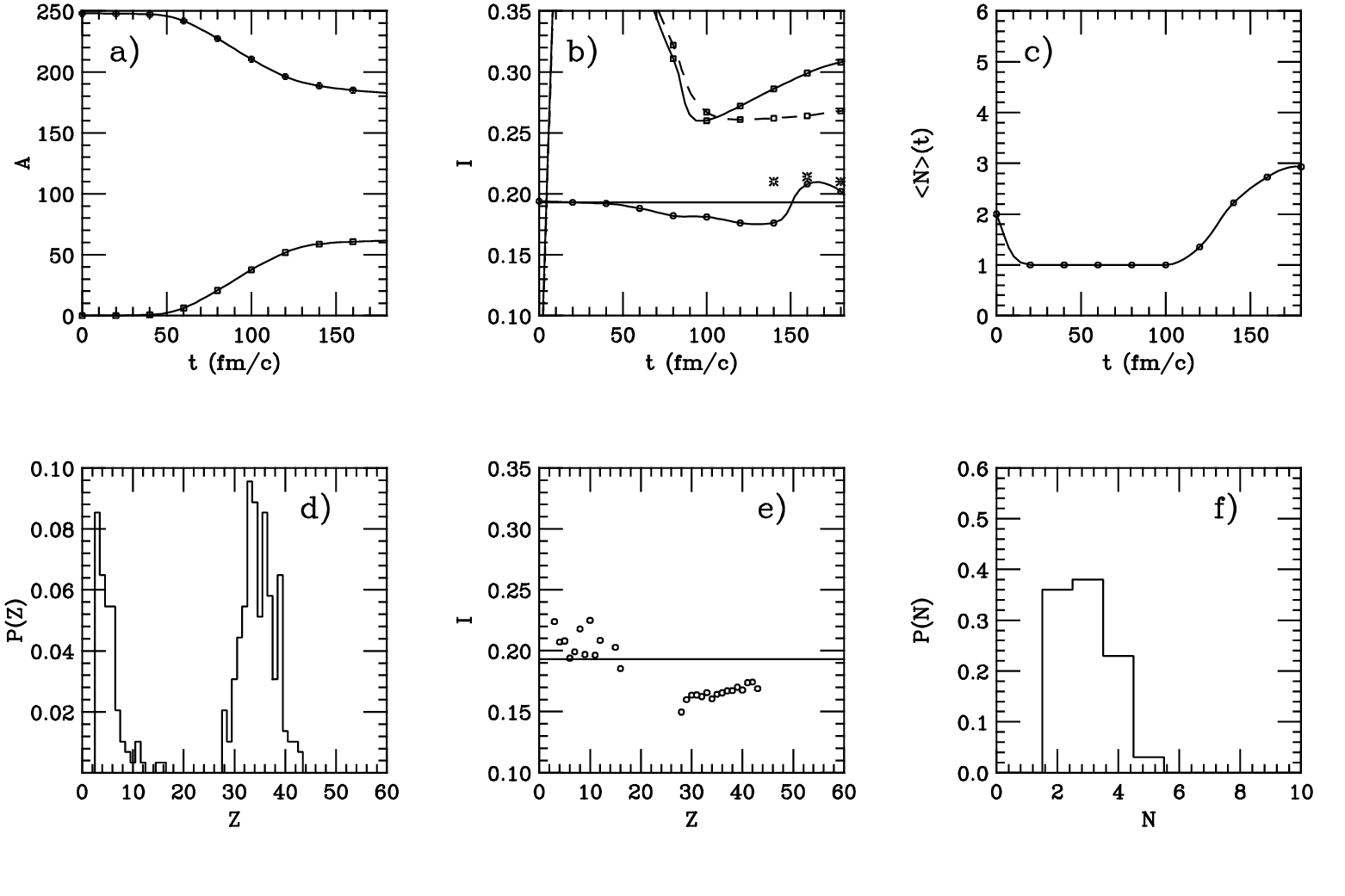,height=5.6truecm}}
\end{center}
{\footnotesize
\begin{quotation}
\noindent
{\bf Fig.4 - } Like Fig.3 for $b=6fm$. 
\end{quotation}}
\end{figure}

\newpage 

We see a neutron dominated prompt particle emission and a second
{\it neutron burst} at the time of fragment formation in
the "central region". The latter is consistent with the dynamical
spinodal mechanism in dilute asymmetric nuclear matter, as
discussed before. The effect is quite reduced for semi-peripheral
collisions (compare Fig.s 3b and 4b) and the $IMF$'s 
($3 \leq Z \leq 12$) produced in 
the neck are more neutron rich (Fig.s 3e and 4e). 
This seems to indicate a different nature of the
fragmentation mechanism in central and neck regions, i.e.
a transition from volume to shape instabilities with different
isospin dynamics. In more peripheral collisions the interaction time
scale is also very reduced (Fig.4c) and this will 
quench the isospin migration.

\begin{figure}
\begin{center}
\mbox{\psfig{file=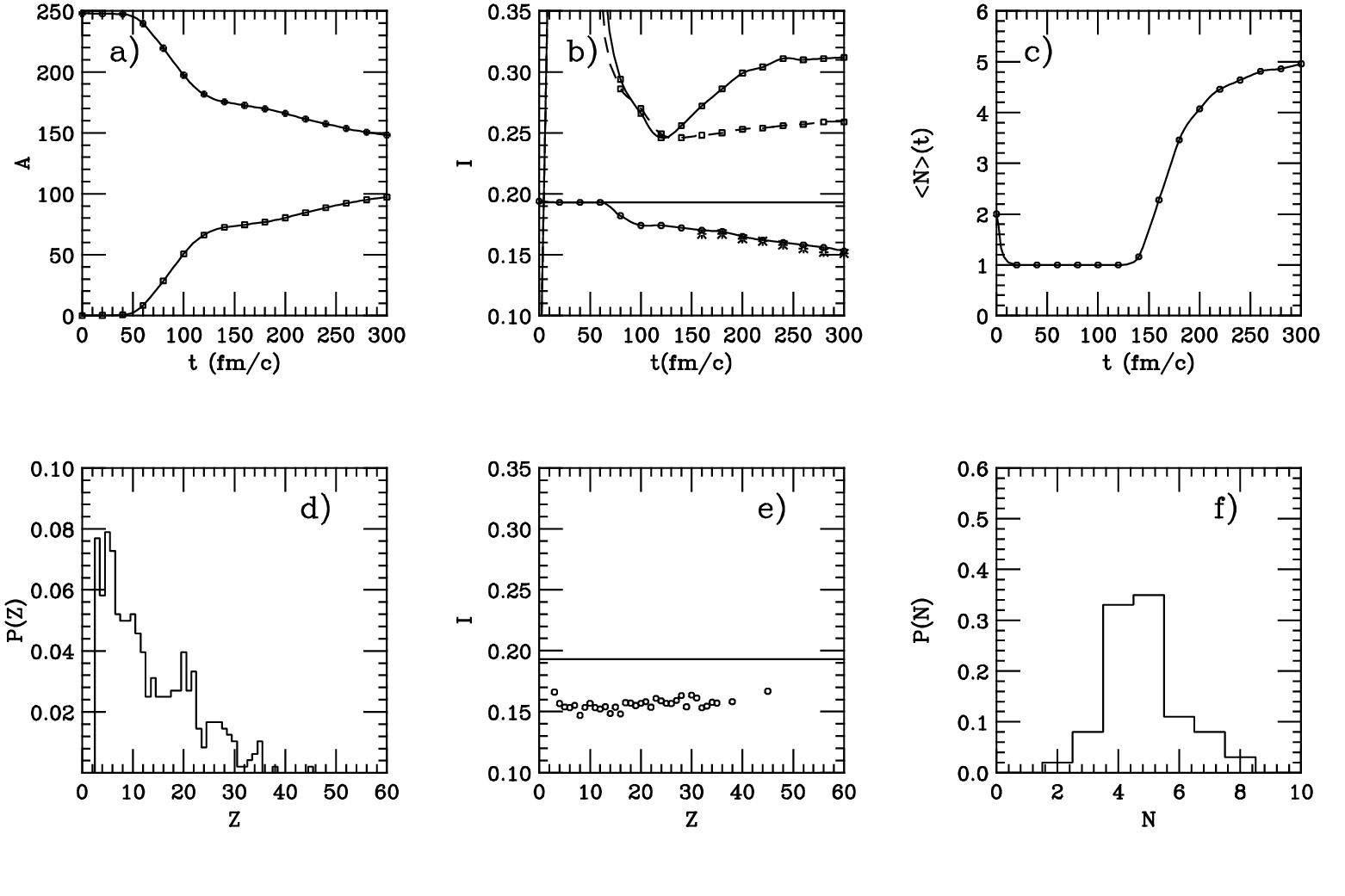,height=5.6truecm}}
\end{center}
{\footnotesize
\begin{quotation}
\noindent
{\bf Fig.5 - } $^{124}Sn+^{124}Sn$ $50AMeV$ $b=2fm$ collisions: time
evolution and freeze-out properties. See text. {\it ASY-SOFT} $EOS$.
\end{quotation}}
\end{figure}

\begin{figure}
\begin{center}
\mbox{\psfig{file=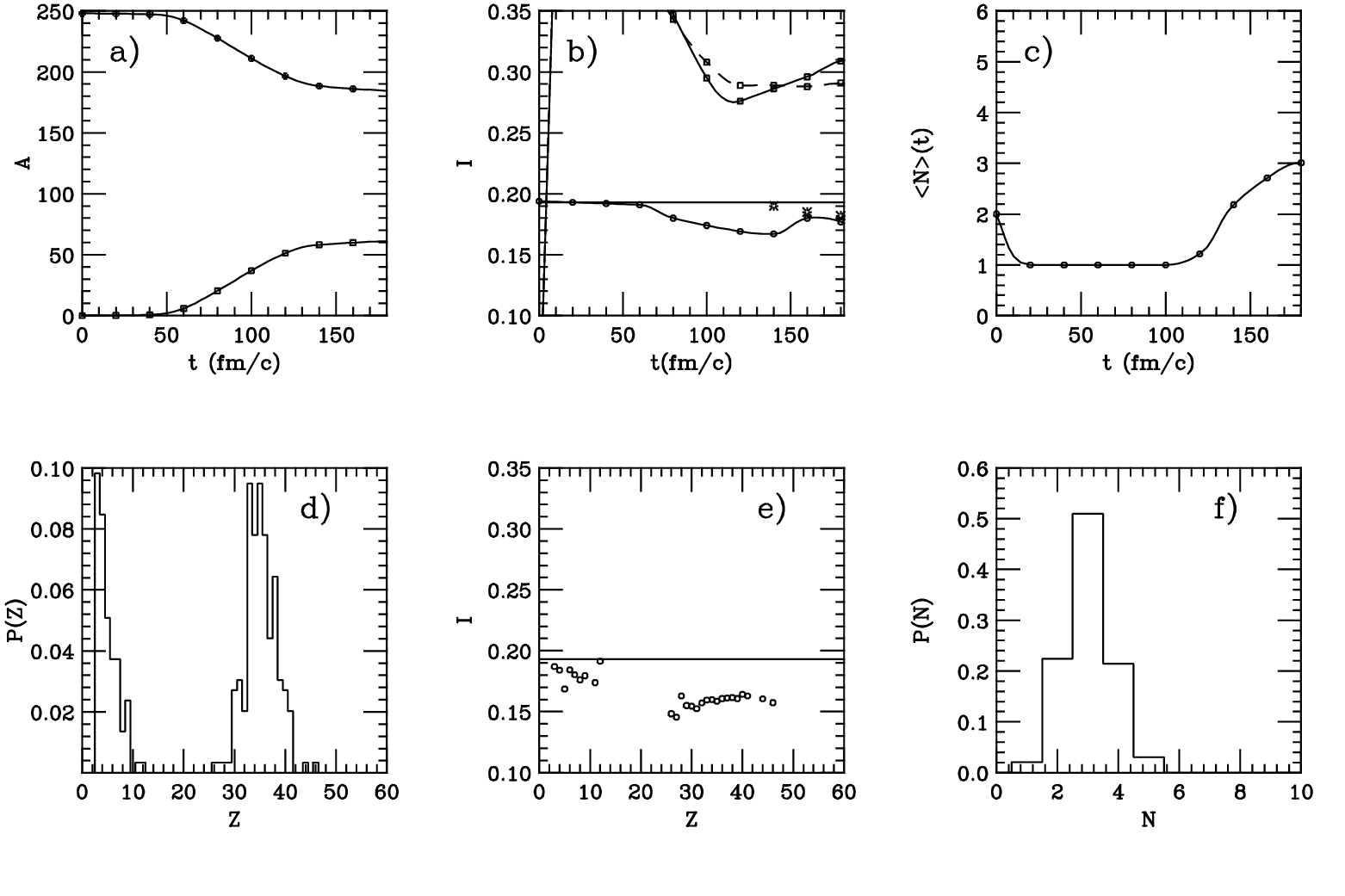,height=5.6truecm}}
\end{center}
{\footnotesize
\begin{quotation}
\noindent
{\bf Fig.6 - } Like Fig.5 for $b=6fm$. 
\end{quotation}}
\end{figure}

In Fig.s 5,6 we have the corresponding results for the {\it asy-soft}
$EOS$. The main qualitative
difference is a larger prompt neutron emission 
(compare the gas asymmetry in Fig.s 3b,4b with the 
corresponding Fig.s 5b,6b) joined to a "slower"
dynamics (see the (c)-plots), as expected from the more attractive 
nature of the
asymmetric $EOS$ \cite{dit98}. 

The interesting point is that 
the effect on fragment production is different for central and
peripheral events. With respect to the {\it asy-stiff} case 
we have a smaller mean $IMF$ multiplicity for central collisions
(Fig.s 3f vs. 5f) and more fragments produced in the neck region
(Fig.s 4f vs. 6f).

Neck instabilities have also the feature of forming clusters on the
"spectator" side \cite{col95} leading to a "fission-like" splitting
of the Projectile-/Target-like fragment. This mechanism has been
clearly observed in accurate kinematical selections
\cite{colin}. In our study of n-rich systems the rate of
such dynamical fission processes is systematically larger
for the asy-soft $EOS$, as expected from the previous discussion. 
This seems to be an quite sensitive observable to look at.

The most important qualitative difference of the neutron poor
case, $^{112}Sn+^{112}Sn$, is a larger prompt proton emission,
in particular for central collisions. The number of protons available
to produce clusters is then reduced with a related smaller
mean $IMF$ multiplicity. Moreover the reduction of
the overall asymmetry is acting in the same direction.
 In Fig.7 we show the average
$IMF$ multiplicity vs. the average charge of the heaviest produced fragment,
which measures the centrality of the collision, for the two symmetry
terms and for the two systems. 
The effect, in agreement with recent data \cite{Mil99}, is more evident
with the {\it asy-stiff} choice, in particular for central collisions
(low $<Z>_{heavy}$).

\begin{figure}
\begin{center}
\mbox{\psfig{file=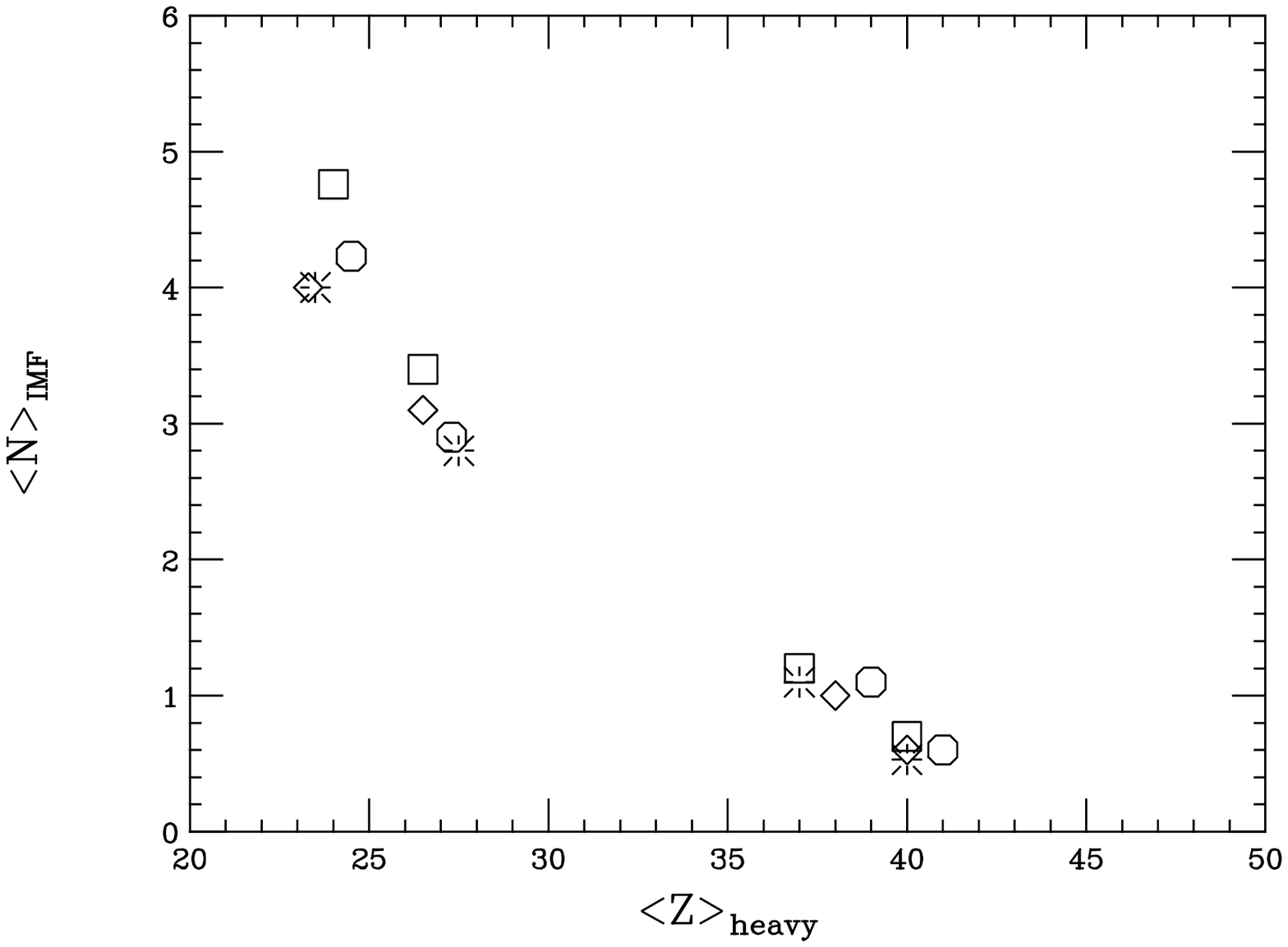,height=4.5truecm}}
\end{center}
{\footnotesize
\begin{quotation}
\noindent
{\bf Fig.7 - } Correlation between mean $IMF$ multiplicity and charge
of the heaviest fragment: {\it squares} n-rich "asy-stiff"; 
{\it diamonds} n-poor "asy-stiff"; {\it circles} n-rich "asy-soft";
{\it stars} n-poor "asy-soft".
\end{quotation}}
\end{figure}

\section{Outlook}
Starting from simulations of reaction dynamics performed
with a new transport code where isospin and fluctuation effects
are suitably accounted for, we have shown that dissipative
heavy ion collisions, and in particular fragment production reactions,
 at medium energies are rather
sensitive to the density dependence of the symmetry contribution to the
nuclear Equation of State. 

Interesting differences appear between isospin effects on central and
neck fragmentation. An indication for a "stiff" symmetry term is
emerging, consistent with transverse flow calculations in the same
energy range \cite{scd99}.

The effects are not large but quite encouraging for similar studies
with radioactive beams.
A possibility is emerging of obtaining in terrestrial accelerator 
laboratories
important information on the symmetry term of large
astrophysical interest. It appears essential to have good
charge asymmetric beams available at intermediate energies
and to perform more exclusive experiments.

\end{document}